%% file: main.tex
\documentclass[conference]{IEEEtran}
\IEEEoverridecommandlockouts

\usepackage{booktabs}
\usepackage{listings}
\usepackage[hang]{footmisc}
\usepackage{amsmath,amssymb,amsfonts}
\usepackage{siunitx}
\usepackage{graphicx}
\usepackage{textcomp}
\usepackage{changes}
\usepackage{csquotes}
\usepackage{xcolor}
\usepackage{hyperref}
\usepackage{setspace}
\usepackage{wrapfig}
\usepackage{float}
\usepackage{algorithm}
\usepackage{algpseudocode}
\usepackage{svg}
\usepackage{arydshln}
\usepackage{cleveref}
\usepackage{subcaption}
\usepackage{enumitem}
\usepackage{titlesec}

\def\BibTeX{{\rm B\kern-.05em{\sc i\kern-.025em b}\kern-.08em
    T\kern-.1667em\lower.7ex\hbox{E}\kern-.125emX}}

\hypersetup{
	colorlinks = true,
	linkcolor = blue,
	urlcolor=blue,
	citecolor=blue
}

\begin{document}

\author{\IEEEauthorblockN{Alexander Berndt}
\IEEEauthorblockA{\textit{Karlsruhe University of Applied Sciences} \\
Karlsruhe, Germany \\
0009-0009-5248-6405}
\and
\IEEEauthorblockN{Zoltán Nochta}
\IEEEauthorblockA{\textit{Karlsruhe University of Applied Sciences}\\
Karlsruhe, Germany \\
0000-0001-9146-8968}
\and
\IEEEauthorblockN{Thomas Bach}
\IEEEauthorblockA{\textit{SAP}\\
Walldorf, Germany \\
0000-0002-9993-2814}
}
\title{The Vocabulary of Flaky Tests in the Context of SAP HANA}

\maketitle

\thispagestyle{plain}
\pagestyle{plain}
\input{abstract}

\footnotetext{Source code available via: \url{https://doi.org/10.5281/zenodo.8107295}\\
978-1-6654-5223-6/23/\$31.00 \copyright 2023 IEEE
}
\begin{IEEEkeywords}
test flakiness, software testing, regression testing, machine learning
\end{IEEEkeywords}
\input{sections/10_lay-abstract}
\input{sections/20_introduction}
\input{sections/30_questions-and-related-work}
\input{sections/40_background}
\input{sections/50_replication-setup}
\input{sections/60_results}
\input{sections/70_conclusion}

\bibliographystyle{IEEEtran}
\bibliography{bibliography}

\end{document}

%% file: abstract.tex
\begin{abstract}
Background. Automated test execution is an important activity to gather information about the quality of a software project. So-called flaky tests, however, negatively affect this process. Such tests fail seemingly at random without changes to the code and thus do not provide a clear signal. Previous work proposed to identify flaky tests based on the source code identifiers in the test code. So far, these approaches have not been evaluated in a large-scale industrial setting.

Aims. We evaluate approaches to identify flaky tests and their root causes based on source code identifiers in the test code in a large-scale industrial project.

Method. First, we replicate previous work by Pinto et al. in the context of SAP HANA. Second, we assess different feature extraction techniques, namely TF-IDF and TF-IDFC-RF. Third, we evaluate CodeBERT and XGBoost as classification models. For a sound comparison, we utilize both the data set from previous work and two data sets from SAP HANA.

Results. Our replication shows similar results on the original data set and on one of the SAP HANA data sets. While the original approach yielded an F1-Score of 0.94 on the original data set and 0.92 on the SAP HANA data set, our extensions achieve F1-Scores of 0.96 and 0.99, respectively. The reliance on external data sources is a common root cause for test flakiness in the context of SAP HANA.

Conclusions. The vocabulary of a large industrial project seems to be slightly different with respect to the exact terms, but the categories for the terms, such as remote dependencies, are similar to previous empirical findings. However, even with rather large F1-Scores, both finding source code identifiers for flakiness and a black box prediction have limited use in practice as the results are not actionable for developers.
\end{abstract}

%% file: sections/10_lay-abstract.tex
\section{Lay Abstract}
We train multiple machine learning models to predict whether a test is flaky or not. As a training input, we use source code identifiers in the test code and historical test results. Furthermore, we identify the features with the highest information gain for test flakiness. Broadly speaking, the keywords that have the highest impact to trigger flakiness.

The models on data from SAP HANA achieve F1-Scores of 75\% and 94\%. The first score for data collected from CI, the second from data collected from a dedicated flakiness experiment. Our results also show trade-offs between techniques with higher investments in training time versus F1-Score gains.

Our results show that \emph{virtual} is the identifier most strongly associated with test flakiness. In context of SAP HANA, \emph{virtual} refers to virtual tables, which represent tables from remote sources. That shows an important practical limitation. We cannot advise developers to stop using \emph{virtual}. Although we know \emph{virtual} has a high chance of leading to flakiness, developers also have to test exactly this functionality. Similarly, if we would use the trained model as a prediction for code changes, the results would not be actionable for developers. They would just get a probability value that their change can lead to flakiness, but no clear instructions what to improve. Therefore, while we can replicate previous results, we conclude that the practical usefulness is limited so far.

%% file: sections/20_introduction.tex
\section{Introduction}
\label{sec:introduction}
With the rise of agile software development and shorter release cycles, companies aim for continuous integration (CI) and continuous delivery (CD)~\cite{rise-ci}. This often requires regression tests to verify that a change does not cause unintended effects~\cite{swebok}. A change can only be automatically integrated into the main code line if all tests show positive results. However, this process is commonly affected by so-called flaky tests. A flaky test occasionally fails seemingly at random, without changes to the executed code. With that, a test may not provide a clear signal and indicate an issue, which does not exist in the code. Thus, flaky tests interfere with the automatic integration of code changes and complicate quality assurance. 

Flaky tests are a common problem in the software industry~\cite{google-flaky, spotify-flakiness, slack-flaky, microsoft-flaky}. 
In the context of large-scale software, where automation has an increasingly important role, test flakiness negatively affects automation and quality assurance. Thus, software companies put a lot of effort into handling flaky tests~\cite{google-flaky-fix, spotify-flakiness, pfs-facebook, survey-flaky-tests, flaky-apple}. 

Previous research shows that a common strategy for dealing with flaky tests is re-executing failed test executions and accepting the test result if one re-execution passes~\cite{barbosa}.  
The CI system for SAP HANA utilizes, among others, the same strategy. When a test case fails, it is re-executed three times. If one of the additional test executions reports a passing result, the test is considered successful~\cite{tbach-newest, survey-flaky-tests}. However, re-executing tests is costly with regard to computational resources. Furthermore, even though the additional test executions can be executed in parallel, the step \emph{re-executing tests} happens after the first execution and is therefore a sequential activity. Such a sequential step negatively affects the overall CI/CD process and increases waiting times for developers considerably. Hence, reducing the number of flaky failures could save computational resources and decrease the turnaround time of integrating code changes. 

Debugging flaky tests is a complex task for developers, because flaky failures might not be reproducible by re-executing the test~\cite{understand-reproducibility}. Therefore, previous work has investigated detecting flaky tests statically, i.e., without relying on (re-)executions. Identifying source code and with that the root causes of test flakiness might help developers to fix and avoid flaky tests in the future~\cite{survey-flaky-tests}.

Recent approaches to detect flaky tests have focused on the use of machine learning algorithms. Assuming that test code yields information about the root cause for the flaky behavior of a test, Pinto et al. propose to detect flaky tests based on the source code identifiers in the test code with promising results~\cite{vocabulary-flaky}. Additionally, by analyzing the employed features for the prediction, they identify source code identifiers associated with flaky tests. These source code identifiers provide information on the underlying root cause for the flakiness. To the best of our knowledge, this approach has not been evaluated in a large-scale industrial context yet. We provide such an evaluation in the context of SAP HANA via replication of previous work~\cite{replications-software-engineering}. Overall, our contributions are:
\begin{enumerate}
    \item A replication of previous work to identify the vocabulary of flaky tests for a large industrial software project.
    \item The evaluation of multiple techniques for predictingthe flakiness of tests. Our comparison includes TF-IDF, TF-IDFC-RF, CodeBERT, and XGBoost.
    \item Discussion of the practical usefulness of the previously mentioned approaches and their findings.
\end{enumerate}

%% file: sections/30_questions-and-related-work.tex
\section{Research Questions and Previous Work}
In this section, we introduce the research questions and relate them to the findings of previous studies.

\textbf{RQ1}: By replicating previous work of Pinto et al.~\cite{vocabulary-flaky}, what F1-Scores do we achieve for the original data set and two data sets collected for SAP HANA? 

Under the assumption that the source code of flaky tests follows certain syntactical patterns, Pinto et al. tried to predict flakiness automatically with the help of Natural Language Processing (NLP) techniques~\cite{vocabulary-flaky}. They evaluated their model on one of the largest data sets for flaky tests obtained by DeFlaker~\cite{deflaker}. The DeFlaker data set consists of over 5000 flaky tests~\cite{deflaker-dataset}. However, as the DeFlaker data set does not contain any non-flaky tests, Pinto et al. re-executed the tests from the different projects in the DeFlaker data set 100 times and labeled tests as non-flaky when they showed consistent results over the 100 executions. In the end, a data set containing approximately 1400 flaky and non-flaky tests was obtained. To analyze the test code, Pinto et al. extracted source code identifiers and split them using their camel-case syntax. After that, they performed common pre-processing steps like stemming and stop word removal and embedded the code identifiers in a bag-of-words. Furthermore, they added the number of Java keywords and the number of lines of code to the feature set to incorporate a proxy for code complexity. Subsequently, the pre-processed data was employed to train several machine learning models, receiving the best F1-Score of 0.95 with a random forest classifier~\cite{random-forest}. We provide an evaluation of Pinto et al.'s study in the context of SAP HANA. 

\textbf{RQ2}: What are the most important features of the prediction model and how do they reflect the most prevalent root causes of test flakiness at SAP HANA?

 Pinto et al. also identified the test code identifiers that are most strongly associated with flakiness. To achieve this, they calculate the information gain from the incorporated features. With that, they could identify the vocabulary of flaky tests in the DeFlaker data set. The resulting vocabulary mainly consisted of words related to remote task execution and event queues. Aside from that, a high number of occurrences of the Java keyword \emph{throw} decreased the likelihood of a flaky test. Pinto et al. conclude that enforcing proper exception handling in the test code can be a measure to prevent test flakiness. 
 In our work, we calculate the information gain of the test code identifiers in two data sets from SAP HANA to identify the most prevalent root causes of test flakiness.
 
\textbf{RQ3}: How do the extensions TF-IDF, TF-IDFC-RF, XGBoost, and CodeBERT compare against the original approach of Pinto et al. in terms of F1-Score?

\textbf{TF-IDF and TF-IDFC-RF for feature embedding}: \emph{Term Frequency - Inverse Document Frequency} (TF-IDF) is a term weighting scheme to represent text as a vector~\cite{idf-definition}. TF-IDF aims at improving Bag-of-Words representations with regard to classification tasks. Therefore, we compare the use of TF-IDF matrices in terms of prediction performance against the original approach using Bag-of-Words. To further advance the weightings of conventional TF-IDF for supervised classification tasks, previous research has proposed a wide range of approaches to perform so-called supervised term weighting~\cite{tf-igm, stw, tf-rf, tf-idf-icf, tf-igm-imp, delta-tfidf}. We evaluate the supervised weighting scheme Term Frequency-Inverse Document Frequency in Classes-Relevance Factor (TF-IDFC-RF) as proposed by Carvalho et al. as this scheme showed the highest results on two benchmarks~\cite{tf-idfc-rf}.

\textbf{XGBoost as a classification model}: Previous research recommends XGBoost in a wide variety of problems~\cite{xgboost}. Thus, we evaluate XGBoost as an alternative implementation.

\textbf{CodeBERT as a classification model}: Fatima et al. used CodeBERT to predict test flakiness based on the source code of the test~\cite{flakify, codebert}. Their evaluation suggests comparable good results on the \emph{FlakeFlagger} data set~\cite{without-rerunning}. Thus, we compare CodeBERT against the approach by Pinto et al.~\cite{vocabulary-flaky}.

%% file: sections/40_background.tex
\section{Background and Data Sets}
In this section, we introduce SAP HANA, the main subject of our study. SAP HANA is a large-scale in-memory database management system with millions of lines of code, mainly written in C++ and developed by SAP~\cite{sap-hana, hana-architecture, hana-paper}.

\subsection{Testing at SAP HANA}
The code of SAP HANA is tested by about one million tests, mostly written in C++ or Python~\cite{tbach-newest}. Roughly speaking, each test can be classified as a unit test or system test. Hereby, unit tests are typically written in C++, whereas system tests are written in Python. In addition to Python code, a large portion of the system tests contain SQL statements to communicate with the SAP HANA instance under test~\cite{tbach-newest}. To train the models in this work, we focus on system tests, as they are more expensive with regard to required hardware, software, and human effort~\cite{tbach-doctor}.

For each test case execution of SAP HANA's test suite, SAP collects metadata and stores it in an internal SAP HANA instance. Among other attributes, this metadata includes the results of each test execution, allowing us to derive flaky labels based on the result history of tests. 

\begin{figure*}
    \centering
    \includegraphics[width=0.6\textwidth]{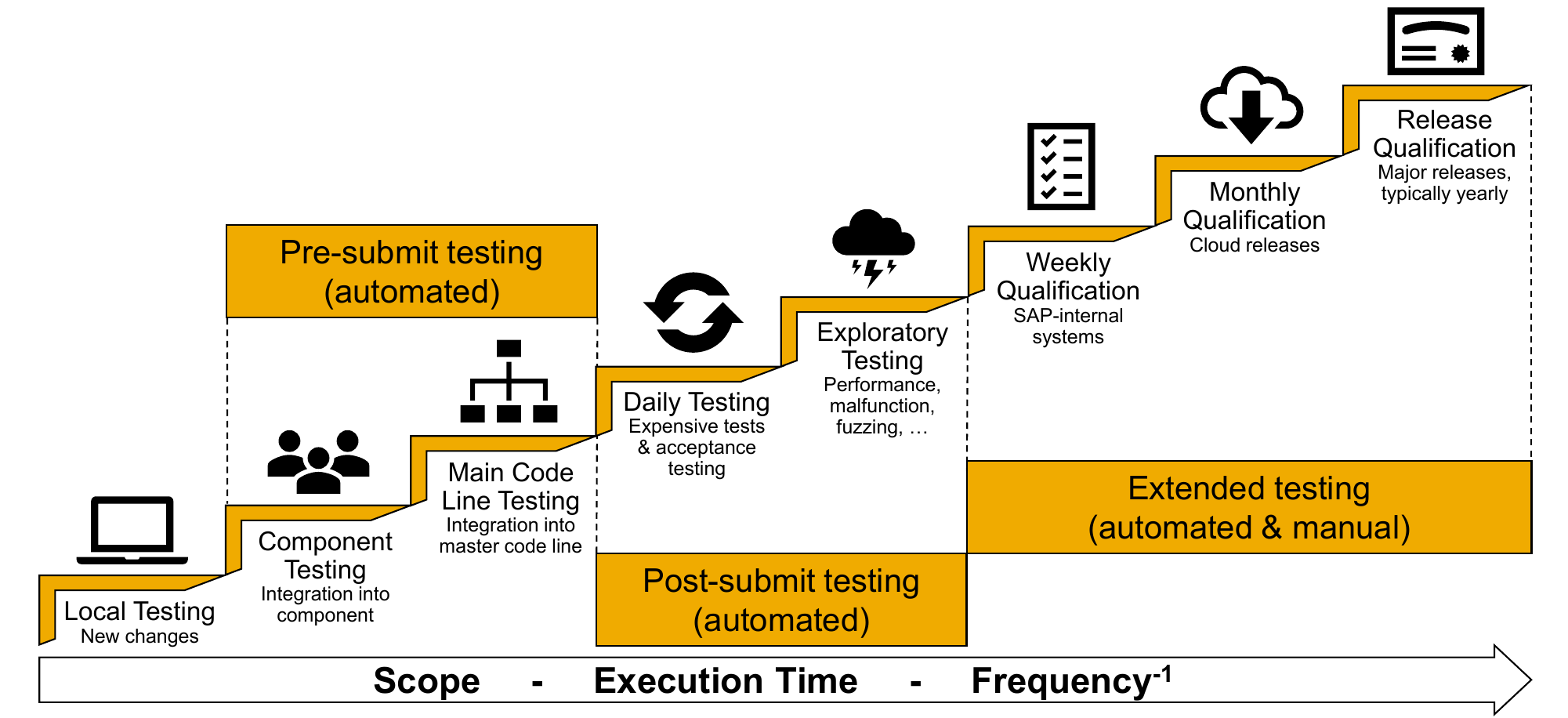}
    \caption{Testing Stages of SAP HANA}
    \label{fig:testingstages}
\end{figure*}

\subsection{Testing Stages}
To assure the quality of SAP HANA, multiple testing stages varying in scope, effort, and frequency have been defined. \Cref{fig:testingstages} provides an overview of the different stages.
SAP collects most of the metadata about test results in three of these stages: pre-submit testing, post-submit testing, and extended testing. For example, within July 2022 alone, more than 800 million test case execution results have been reported. This amount of data enabled us to collect two flaky test data sets from SAP HANA: \emph{Mass Test Execution 2020}  (MTE20) and \emph{Flaky Test 2021} (FT21).

\subsection{MTE20}
The so-called \emph{\enquote{Mass Test Execution}} was conducted in 2020 and refers to a research project in which a subset of SAP HANA's tests was executed \num{100} times against the same build of the code. Hereby, all test results were persisted and thus were available for our work. Based on these test results, \num{35000} test cases could be labeled as flaky or not.

For the labeling, we calculated the average result $\overline{x}_s$ of each test case $s$ over the 100 executions. Hereby, 0 referred to a passing and 1 to a failing execution. 
A test case was labeled as flaky, when $\overline{x}_s \not\in \{0,1\}$.

\subsection{FT21}
The purpose of the FT21 data set was to examine whether our adopted approach is also applicable to data from SAP HANA's daily business. Therefore, we collected the data from test results from the CI system of SAP HANA in 2021. As mentioned before, failing tests in CI runs are re-executed three times for the same code. Similar to our labeling for MTE20, we labeled a test as flaky, when it yielded different results throughout these four executions.

However, since SAP HANA's tests are executed in a permanently evolving environment, test failures can be caused by so-called \enquote{global issues}, i.e., problems that were not caused by the test code. For example, in July 2021, global issues affected roughly 10\% of the builds in SAP HANA's CI pipeline. In many cases, such global issues caused flaky labels even though the failure was not related to the actual test code. However, in the context of our work, these labels could be considered noise and interfere with the training of our model~\cite{class-noise}. Hence, we applied a heuristic approach to filter failures which were caused by global issues. 

Based on the assumption that global issues affect multiple tests simultaneously, the following filter was applied: 
Let $s_i$ be the number of passing test cases and $n_i$ the total number of executed tests for build $i$ of SAP HANA.

To filter out builds that were affected by a global issue, we only consider tests from builds that fulfill both of the following conditions:
\begin{enumerate}[label=(\roman*)]
        \item $\frac{s_i}{n_i} > 0.99$ 
        \item $n_i > 1000$
\end{enumerate} 
That is, we took only test results from builds with at least 1,000 executed tests into account, of which at least 99\% have been successful. By doing so, we obtained labels for approximately 15,000 test cases. 

\subsection{MSR4Flakiness}
To compare the results of our replication setup and its extensions against the results from the original study, we employed the MSR4Flakiness data set in addition to the two data sets from SAP HANA.
Pinto et al. collected the data for the MSR4Flakiness data set for their work~\cite{msr4flakiness, vocabulary-flaky}.

\subsection{Exploratory Data Analysis}
To amplify our intuition about the available data from SAP HANA, we first conducted a manual examination of the test code with regard to test flakiness~\cite{google-eda}. Therefore, we assessed commits, which were supposed to fix flaky labeled bugs. To compare the results against findings from previous literature, commits were divided into the categories proposed by Luo et al.~\cite{empirical} as shown in \Cref{tab:examine-flaky-commits}. Besides, we added the category \emph{Fixed Timeout}, because increasing the maximum duration of a test was one of the most prevalent fixes for flaky behavior in the context of SAP HANA. Hereby, the maximum duration refers to the maximum execution time of a test before it is canceled. \Cref{tab:examine-flaky-commits} shows the results of the analysis.

\input{tables/examine-flaky-commits.tex}

The three most prevalent categories for test flakiness in the assessed commits are \emph{Fixed Timeout}, \emph{Concurrency} and \emph{Async Wait}. These findings are in line with the results provided by previous work. Luo et al. report the same categories~\cite{empirical}. A noticeable difference to previous findings are large number of commits in the category \emph{Hard to classify}. However, this gap can be explained by differences in the methodology. Luo et al. time-boxed their approach to 2 hours per commit, whereas we used an overall time-box of 8h, which equals to roughly 10 minutes per commit. That means we spent less time per commit for classification and our project might be larger and therefore more complex to find a definitive classification.

%% file: tables/examine-flaky-commits.tex
\begin{table}
  \centering
  \caption{Root Causes of Flaky Tests}
  \begin{tabular}{l r r}
      \textbf{Root Cause} & \textbf{Amount} & \textbf{Percentage} \\
      \cmidrule{1-3}
      Fixed Timeout & 8 & 17\%\\
      Concurrency & 7 & 15\%\\
      Async Wait & 6 & 11\%\\
      Test Order Dependency & 6 & 11\%\\
      Unordered Collections & 3 & 6\%\\
      Randomness & 2 & 4\%\\
      Resource Leak & 1 & 2\%\\
      Time & 1 & 2\%\\
      Hard to Classify & 13 & 28\%\\
      Network & 0 & 0\%\\
      IO & 0 & 0\%\\
      \cmidrule{1-3}
      Total & 47\\
      \end{tabular}
    \label{tab:examine-flaky-commits}
\end{table}
\raggedbottom

%% file: sections/50_replication-setup.tex
\section{Replication Setup}
As described in \Cref{sec:introduction}, we conduct a conceptual replication of a previous study~\cite{replication-conceptual, vocabulary-flaky}. This section provides an overview of the steps conducted to implement our replication.

\subsection{Data Resampling}
A common problem with data related to test flakiness is the imbalance between non-flaky and flaky samples~\cite{without-rerunning}. Typically, data sets contain a higher number of non-flaky samples~\cite{neighbor-flakiness}. This problem applies to our data sets as well. \Cref{tab:class-distribution} displays the class distribution of the data sets, showing that all of them contain more samples from the non-flaky class. This leads to the conclusion that all of them are imbalanced. 

Using an imbalanced data set to train a machine learning classifier can be problematic as the classifier might be skewed towards the majority class. That is, in the context of the data sets for this work, the classifier could overestimate the probability that a sample is non-flaky. 
\input{tables/class-distributions.tex}

A common technique to handle imbalanced data is \emph{undersampling}~\cite{undersampling-paper}. Hereby, the data set is balanced out by keeping all the samples from the minority class, whilst decreasing the size of the majority class by randomly dropping samples. For this work, we implemented undersampling with \emph{RandomUnderSampler}~\cite{undersampler-docs} provided by \emph{imbalanced-learn}~\cite{undersampler-docs}. We employed undersampling as the first step of our pipeline, as it reduces the total amount of samples to be processed, thus saving computational resources for future steps.

\subsection{Tokenization}
\label{sec:tokenization}
Manning et al. define tokenization as the task to chop a document into pieces~\cite{tokenization-definition}. They call the resulting pieces tokens and state that \enquote{tokens are often loosely referred to as terms or words}~\cite{tokenization-definition}. Tokenization can also involve the removal of unwanted characters, for example, punctuation. For this work, we processed tokens depending on their type and content as defined by the \emph{token} library~\cite{token-docs}:
\begin{enumerate}
    \item \textbf{Keywords}: 
    As Python keywords should be treated differently than source code identifiers in a later step of the pipeline, we identified and marked them during tokenization.
    We also add the token \emph{self} to the list of keywords. Although \emph{self} is not a keyword in Python, it is conventionally used.
     \item \textbf{Numbers}: 
    Previous research has shown that one of the most prevalent categories for test flakiness is \emph{Async Wait}~\cite{empirical}. For example, in the context of SAP HANA, we found flaky tests, which use \emph{time.sleep(n)} to wait \emph{n} milliseconds for an external task to finish. To grasp all these occurrences in one token, numbers were masked with the \emph{\#NUM\#} token.
    \item \textbf{Names}: 
    Tokens of type \emph{NAME} were added to the resulting list of tokens. 
    \item \textbf{String}:
    Tokens of type \emph{STRING} were tokenized under the use of \emph{word\_tokenize} as provided by \emph{nltk}~\cite{nltk-docs}. 
\end{enumerate}

\begin{figure}
    \centering
    \includegraphics[width=0.5\textwidth]{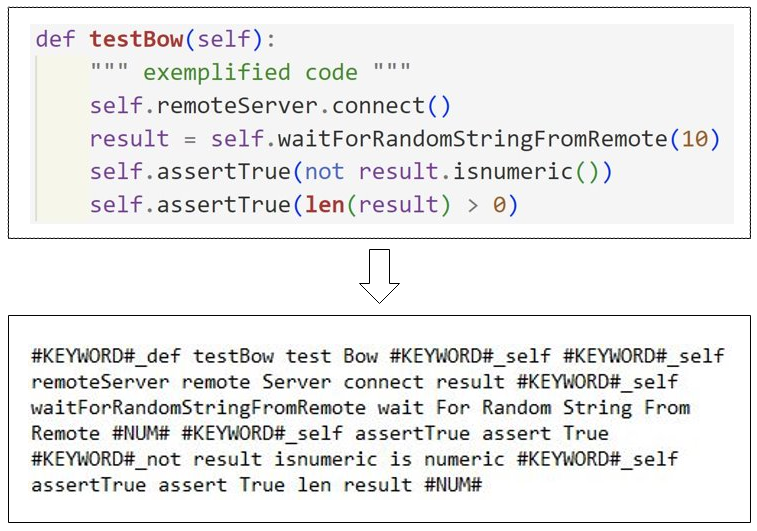}
    \caption{Example of Tokenization}
    \label{fig:tokenization}
\end{figure}

\Cref{fig:tokenization} shows the resulting tokens for an exemplified test after we apply the described steps. In contrast to the study by Pinto et al., we did not apply stemming and stop word removal, as the results of Pinto et al. showed that these steps did not have any effect on the resulting model.

\subsection{Identifier Splitting}
The study by Pinto et al. has shown that identifier splitting can have a positive impact on the performance of a vocabulary-based classifier for test flakiness prediction~\cite{vocabulary-flaky}. Identifier splitting refers to the task of tokenizing source code identifiers. A common way to achieve this is to split source code identifiers based on their case style~\cite{identifier-splitting-paper}. For example, the identifier \emph{waitForSomething}, which is written in \emph{\enquote{camelCase}}, can be chopped up into the tokens \emph{wait}, \emph{For}, and \emph{Something}. 

In the original study, Pinto et al. split the identifiers based on their camel-case syntax. However, in the context of SAP HANA, we found that the test code contains different casing styles. To split source code identifiers based on different casing styles, Hucka et al. introduced the package \emph{Spiral}~\cite{spiral}. The package Spiral implements a range of splitting algorithms such as the Ronin algorithm. The Ronin algorithm was already employed in the context of SAP HANA in a previous study~\cite{gabin}. Thus, we considered it suitable as a tokenizer to split identifiers for this work. \Cref{tab:spiral-result} shows the resulting split identifiers from \Cref{fig:tokenization} under the use of Spiral and Ronin. As proposed by Pinto et al.~\cite{vocabulary-flaky}, all the split tokens together with the original identifier were added to the result set. 
\input{tables/ronin.tex} 

\subsection{Feature Extraction}
To enable a machine learning classifier to learn from data, the characteristics of the data have to be captured in a set of vectors~\cite{feature-extraction}. We evaluate three approaches to achieve this: Bag-of-Words, TF-IDF, and TF-IDFC-RF.

As proposed by Camara et al.~\cite{vocabulary-flaky-replication}, we employ the \emph{CountVectorizer} class from \emph{Scikit-Learn} to implement Bag-of-Words. For TF-IDF, the respective \emph{TfIdfVectorizer} class was used.
The implementation of TF-IDFC-RF in this work is based on the Python implementation provided by Carvalho et al.~\cite{tf-idfc-rf, tf-idfc-rf-github}. In this implementation, the \emph{Tfidfcvectorizer} class is provided, which implements fit and transform methodology similar to classes from \emph{Scikit-Learn}. 

\subsection{Evaluation}
To train and evaluate a machine learning classifier, it is common practice to split the available data into separate training and test data~\cite{vocabulary-flaky, without-rerunning, flakify}. To achieve valid evaluation results, it is important to avoid data leakage, i.e. utilizing the test data during the learning setup. For example, Arp et al. suggest that a common mistake in the context of NLP tasks is to compute TF-IDF weightings on the complete data set instead of deriving them only from the training data~\cite{dos-donts}. To avoid such mistakes, we split the data for this work before embedding the code into feature vectors.

To evaluate the classifiers for this work, we conduct k-fold validation with $k = 5$ folds under the use of the \emph{KFold} class provided by \emph{Scikit-Learn}~\cite{kfold-docs}.

\subsection{CodeBERT}
In contrast to the other evaluated extensions, we set up a different preprocessing and evaluation pipeline for CodeBERT. To predict whether a test is flaky or not, CodeBERT does not require prior feature extraction but captures potential patterns in the source code automatically. With regard to evaluation, we did not use k-fold cross-validation, because of the increased computational cost to train CodeBERT. Instead, we employed a train-test-split, using 80\% of the data for training and 20\% as a test set.

%% file: tables/class-distributions.tex
\begin{table}
  \begin{center}
  \caption{Distribution of Classes Across Data Sets}
    \begin{tabular}{l r r r}
      \textbf{Data Set} & \textbf{Total} & \textbf{Non-Flaky} & \textbf{Flaky}\\
      \cmidrule{1-4}
      FT21 & 14,703 & 13,566 & 1,137 \\
      MTE20 & 36,963 & 34,311 & 652\\
      MSR4Flakiness & 45,381 & 44,011 & 1,370
      \end{tabular}
    \label{tab:class-distribution}
  \end{center}
\end{table}

%% file: tables/ronin.tex
\begin{table}
  \begin{center}
    \caption{Results of Ronin Algorithm}
    \begin{tabular}{l l}
      \textbf{Input String} & \textbf{Resulting Tokens}\\
      \cmidrule{1-2}
      testBow & ['test', 'Bow']\\
      remoteServer & ['remote', 'Server']\\
      assertTrue & ['assert', 'True']
      \end{tabular}
    \label{tab:spiral-result}
  \end{center}
\end{table}

%% file: sections/60_results.tex
\section{Results}
In this section, we answer the research questions. 

\subsection{Research Question 1: Flaky Test Detection Benchmark}
We replicate the setup of vocabulary-based models as proposed by previous studies and evaluate the replication on the two data sets from SAP HANA~\cite{vocabulary-flaky, vocabulary-flaky-replication, vocabulary-flaky-replication2}. We verify the implemented pipeline and models for this work by comparing our results against previous benchmarks on the MSR4flakiness data set~\cite{msr4flakiness}. For the evaluation, we use 5-fold cross-validation, i.e., we trained and evaluated every model five times. We obtain the final result by rounding the results of the five runs to the second decimal and calculating the arithmetic mean. \Cref{tab:results-replication-hana,tab:results-replication} show the respective results.
\input{tables/results-replication.tex}
The replication yields similar results on the MSR4flakiness benchmark compared to previous studies~\cite{vocabulary-flaky, vocabulary-flaky-replication}.
The results for the two data sets from SAP HANA differ. While the result on the MTE20 data set is similar to the result on the MSR4Flakiness data set, the result on the FT21 data set is worse. However, this difference may be explained by the findings of Haben et al.~\cite{vocabulary-flaky-replication2}. In their replication of Pinto et al.'s study, they found that vocabulary-based models suffer from a time-sensitive evaluation. That is, the performance of their models dropped when they used code from older revisions for training a model to predict flakiness for code from newer revisions. Likewise, the FT21 data set contains commits from a time frame of one year, i.e., the data is based on a range of revisions. In contrast, the data for the MTE20 data set contains only one revision.

Another possible explanation is that we collected the FT21 data set from test results of SAP HANA's CI pipeline, while the data in the MTE data set was purposefully generated to gain insights about test flakiness. Thus, the FT21 data set could be more strongly affected by noise than the MTE20 data set.

\textbf{Conclusion}: Our replication shows an F1-Score of 94\% for MTE20 and 75\% for FT21. 
\input{tables/results-rq4.tex}

\begin{figure}
    \centering
    \includegraphics[width=\columnwidth]{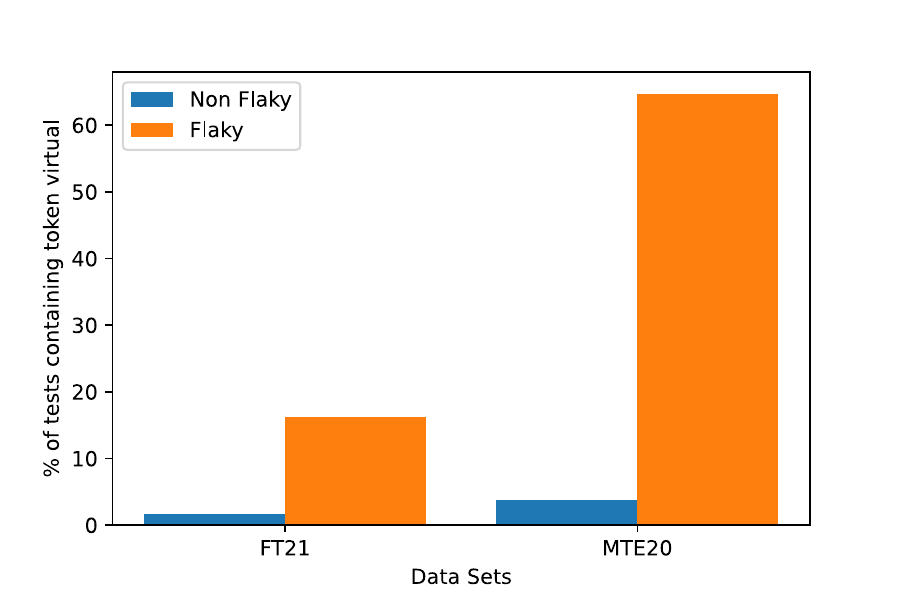}
    \caption{Occurrences of \emph{virtual}}
    \label{fig:virtual-comparison}
\end{figure}

\subsection{Research Question 2: Detect Root Causes}
The second research question aims at finding the root causes of test flakiness in context of SAP HANA. To achieve this, we calculate the information gain from the different features as proposed by Camara et al. \cite{vocabulary-flaky-replication}. \Cref{tab:features-ft21,tab:features-mte20} show the 10 most important features for the FT21 and MTE20 data set, respectively.
Comparing \Cref{tab:features-ft21} and \Cref{tab:features-mte20}, the best features for both data sets are similar. Although in a different order, 4 out of the 10 best features are present for both data sets. For both data sets, the most important feature is \emph{virtual}. Looking at the underlying code, the most common usage of the token \emph{virtual} in SAP HANA's tests is in the context of so-called virtual tables. Virtual tables provide access to tables on a remote source~\cite{hana-documentation}. This finding is in line with previous research, which points out that the reliance on remote sources is a typical root cause for test flakiness~\cite{empirical, vocabulary-flaky, survey-flaky-tests}.

To underline the extent of this finding, we count the number of flaky samples containing \emph{virtual}.
As shown in \Cref{fig:virtual-comparison}, \emph{virtual} appears in approximately 65\% of the flaky samples, but only in 4\% of the non-flaky samples from the MTE20 data set. In absolute numbers, 422 out of 652 flaky samples in the MTE20 data set contain the token \emph{virtual}. However, the token is only contained in 24 non-flaky samples. Predicting a test to be flaky if \emph{virtual} is present would result in a precision of approximately 95\%. 
In the FT21 data set, \emph{virtual} appears in approximately 16\% (184 out of 1.137) of the flaky samples and 1.6\% (18 out of 1.137) of the non-flaky samples. A prediction based solely on the appearance of \emph{virtual} would result in a precision of approximately 91\%.

Since the vocabulary of flaky samples appears similar in both data sets, we can conclude that both data sets contain a similar set of flaky samples. Although the labeling strategy and the examined time frame of the two data sets are completely different, they share 121 flaky labeled test cases. The vast majority of these tests verify the data federation functionality of SAP HANA, which is closely related to virtual tables.

In addition to \emph{virtual}, the two data sets share three additional top predictors, namely \emph{doquery2}, \emph{sameresults}, and \emph{sameexplain}.
Looking at the occurrences of these tokens, it appears that the three tokens are connected. The token \emph{doquery2} refers to a function which implements the functionality to execute a query on the SAP HANA instance under test. The tokens \emph{sameresults} and \emph{sameexplain} refer to parameters of \emph{doquery2}.  As the \emph{doquery2} function is defined in a superclass of the test cases for data federation, these tokens are also interconnected to the token \emph{virtual}.
In fact, all top predictors of the MTE20 data set arise from the context of data federation.

Regarding the FT21 data set, we did not examine every top predictor due to their low level of information gain.

\textbf{Conclusion}: The token \emph{virtual} is the feature, which is most associated with flakiness in both data sets from SAP. Similarly, remote dependencies are a common root cause for flaky tests in both data sets from SAP.
\input{tables/results-rq2.tex}
\input{tables/results-rq3.tex}

\subsection{Research Question 3: Evaluate Extensions}
For the third research question, we evaluate TF-IDF and TF-IDFC as alternative term weighting schemes. In addition, we evaluate XGBoost as an alternative classification model.\Cref{tab:tablerq2-1,tab:tablerq2-3} show the results for the evaluation runs. The values in the table depict the averaged F1-Scores from the 5-fold cross-validation.  
The results in \Cref{tab:tablerq2-1,tab:tablerq2-3} show that the use of term weighting schemes instead of Bag-of-Words yields similar or better performance for every classifier on every data set. However, using supervised term weightings (TF-IDFC-RF) does not result in a better performance when compared to conventional TF-IDF. 

Regarding model selection, XGBoost did not yield improved results when compared to Random Forest. In contrast, Random Forest outperformed XGBoost on all three data sets.

Finally, we fine-tune and evaluate CodeBERT on the three available data sets. As shown in \Cref{tab:results-rq3}, CodeBERT outperforms the previous approaches on every data set, providing state-of-the-art results on the MSR4flakiness data set.

However, while CodeBERT increases the previous prediction performance, it comes with the drawback of higher computational cost and runtime. Training and evaluating CodeBERT took over 10 times longer compared to random forest.

\textbf{Conclusion}:
Using term weighting schemes improved the F1-Score with the largest improvement on the FT21 data set from 75\% with Bag-of-Words to 78\% with the two term weighting schemes TF-IDF and TF-IDFC-RF.
CodeBERT has higher F1-Scores on each examined data set. The MTE20 data set shows the largest increase from 93\% F1-Score to 99\% F1-Score. However, the improvement in prediction performance requires at least a factor 10 increase in training time (10 min versus 2 hours). The resulting model requires over 100 times more disk space (4 versus 400 megabyte).

%% file: tables/results-replication.tex
\begin{table}
    \begin{center}
        \caption{Results Replication SAP HANA}
        \label{tab:results-replication-hana}
        \begin{tabular}{l r r r} 
          \textbf{Data Set} & \textbf{Precision} & \textbf{Recall} & \textbf{F1-Score}\\
          \cmidrule{1-4}
          MSR4Flakiness       & 0.94 & 0.94 & 0.94\\
          FT21      &  0.76 & 0.76 & 0.75\\
          MTE20       & 0.97 & 0.87 & 0.92\\
      \end{tabular}
    \end{center}
\end{table}
\begin{table}
    \begin{center}
        \caption{Comparison to Original Approach}
        \label{tab:results-replication}
        \begin{tabular}{l r r r} 
          \textbf{Approach} & \textbf{Precision} & \textbf{Recall} & \textbf{F1-Score}\\
          \cmidrule{1-4}
          Original         & 0.99 & 0.91 & 0.95\\
          Replication      &  0.94 & 0.94 & 0.94\\
       \end{tabular}
    \end{center}
\end{table}

%% file: tables/results-rq4.tex
\begin{table}
    \caption{Top Features by Information Gain for FT21}
    \label{tab:features-ft21}
    \begin{center}
        \begin{tabular}{l r} 
        \textbf{Feature} & \textbf{Information Gain}\\
        \cmidrule{1-2}
        virtual & 0.039\\
        esh & 0.029\\
        expectedresult & 0.025\\
        \_runtest & 0.023\\
        doquery2 & 0.022\\
        doexplain & 0.020\\
        virtual\_product & 0.020\\
        uvalue & 0.019\\
        doexecute & 0.019\\
        sameresults & 0.019\\
        \end{tabular}
    \end{center}
\end{table}
\begin{table}
    \caption{Top Features by Information Gain for MTE20}
    \label{tab:features-mte20}
    \begin{center}
        \begin{tabular}{l r} 
        \textbf{Feature} & \textbf{Information Gain}\\
        \cmidrule{1-2}
        virtual & 0.239\\
        doquery2 & 0.187\\
        explain & 0.169\\
        do & 0.165\\
        doexecute & 0.158\\
        sameresults & 0.156\\
        sameexplain & 0.121\\
        same & 0.120\\
        dxchg & 0.113\\
        doexplaindxchg & 0.113\\
        \end{tabular}
    \end{center}
    \label{tab:results-rq4}
\end{table}

%% file: tables/results-rq2.tex
 \begin{table}
      \centering
        \begin{tabular}{l r r r}
          \textbf{Model} & \textbf{BoW} & \textbf{TF-IDF} & \textbf{TF-IDFC-RF}\\
          \cmidrule{1-4}
          Random Forest & \textbf{0.94} & \textbf{0.94} & \textbf{0.94}\\
          XGBoost       & 0.92 & \textbf{0.94} & \textbf{0.94}\\
          Randomized    & 0.48 & 0.48 & 0.48\\
          Only True & 0.34 & 0.34 & 0.34 \\
          Only False & 0.32 & 0.32 & 0.32 \\
          \end{tabular}
        \caption{Resulting F1-Scores for MSR4Flakiness}
        \label{tab:tablerq2-1}
  \end{table}
  \par\bigskip
    \begin{table}
        \centering
        \begin{tabular}{l r r r} \textbf{Model} & \textbf{BoW} & \textbf{TF-IDF} & \textbf{TF-IDFC-RF}\\
          \cmidrule{1-4}
          Random Forest & 0.75 & \textbf{0.78} & \textbf{0.78}\\
          XGBoost       & 0.71 & \textbf{0.75} & \textbf{0.75}\\
          Randomized    & 0.48 & 0.48 & 0.48\\
          Only True & 0.34 & 0.34 & 0.34 \\
          Only False & 0.32 & 0.32 & 0.32 \\
          \end{tabular}
        \caption{Resulting F1-Scores for FT21}
        \label{tab:tablerq2-2}
  \end{table}
  \par\bigskip
  \begin{table}
    \centering
        \begin{tabular}{l r r r}
          \textbf{Model} & \textbf{BoW} & \textbf{TF-IDF} & \textbf{TF-IDFC-RF}\\
          \cmidrule{1-4}
          Random Forest & 0.92 & \textbf{0.93} & \textbf{0.93}\\
          XGBoost       & 0.91 & \textbf{0.92} & \textbf{0.92}\\
          Randomized    & 0.48 & 0.48 & 0.48\\
          Only True & 0.34 & 0.34 & 0.34 \\
          Only False & 0.32 & 0.32 & 0.32 \\
          \end{tabular}
        \caption{Resulting F1-Scores for MTE20}
        \label{tab:tablerq2-3}
  \end{table}

%% file: tables/results-rq3.tex
\begin{table}
  \begin{center}
    \begin{tabular}{l r r} 
        \textbf{Data Set} & \textbf{CodeBERT} & \textbf{Previous Best}\\
      \cmidrule{1-3}
      MSR4Flakiness & \textbf{0.96} & 0.94\\
      FT21 & \textbf{0.82} & 0.78\\
      MTE20 & \textbf{0.99} & 0.93\\
      \end{tabular}
    \caption{Resulting F1-Scores of CodeBERT}
    \label{tab:results-rq3}
  \end{center}
\end{table}

%% file: sections/70_conclusion.tex
\section{Threats to validity}

\subsection{Internal Validity} 
A possible threat to the internal validity of the conducted study lies in our data sets. Since we derived the labels from empirical observations, they could be affected by noise and might introduce bias to the results of this work. Since every test has a certain probability to fail occasionally, labeling tests as non-flaky, because they did not fail in a given time frame, could be considered a threat to validity. We mitigate this threat by employing two independent data sets from the same project.

Another possible threat to the internal validity are potential errors in the setup of the replication study. To minimize this threat, we compared our implementation against results from previous research~\cite{vocabulary-flaky}. The comparison shows similar results.

\subsection{External Validity} 
The main threat to the external validity of this work is the unique context of SAP HANA. Although previous research evaluates a similar approach on other data sets, such approaches were not tested on a large-scale software project like SAP HANA. While the results of this work are in line with previous findings, we still consider it valuable to apply this study to further industrial and large projects. Thereby, we  encourage studies on other larger data sets to further evaluate the generalizability of the presented approach.

\subsection{Construct Validity}
The main threat to the construct validity is our implementation to retrieve the test code of SAP HANA. We identify the test code for test results via an automated approach. Due to the complexity of the test suite and test configurations, this automated mapping might not be accurate. We manually verified several cases together with engineers from SAP.

\section{Conclusion}
We conceptually replicated previous work to detect flaky tests based on source code identifiers in the test code in a large-scale industrial project, namely SAP HANA. We compared the results of our replication against previous work on the original data set. With an F1-Score of 94\%, our replication yields similar results compared to the original study (95\%).

Our replication shows an F1-Score of 92\% on the MTE20 data set of SAP HANA. We conclude that vocabulary-based models can detect flaky tests in a large-scale industrial setting.

Furthermore, we apply Pinto et al.'s approach to detect the root causes of flaky tests to SAP HANA's data sets. The presence of the token \emph{virtual} in a test was the best predictor to distinguish between flaky and non-flaky tests. This leads to the conclusion that a common root cause for test flakiness at SAP HANA are virtual tables, which allow to access data from remote sources. This finding is in line with previous research, which considers the reliance on external resources the most common root cause for test flakiness~\cite{empirical,bugs-in-test-code,understanding-flaky-tests,lifecycle-flaky}.

Finally, we evaluate various extensions to the original approach with regard to prediction performance. Hereby, we achieve the best result using CodeBERT with an F1-Score of 99\% on the MTE20 data set. However, the use of CodeBERT comes with the drawback of increased computational cost.

From a practical point of view, the insights from this work are two-fold. On the one hand, we found that the trained models achieved better performance on MTE20. Therefore, we conclude that it is valuable to re-execute tests with the purpose of test flakiness examination. On the other hand, even though the trained models in this study yielded decent performance, we did not employ them in our developing process for two reasons. First, 
 we expected the accuracy of the models to decay over time due to the ever-evolving code of SAP HANA. Second, testing the functionality of virtual tables is essential.